\documentclass[12pt,notitlepage,amsmath,amssymb,pra,twocolumn,floats,aps]{revtex4-2}

\newcommand{\fH}{\mathcal{H}}

\newcommand{\fL}{\mathcal{L}}

\newcommand{\tR}{\mathrm{R}}

\newcommand{\tI}{\mathrm{I}}
\newcommand{\tT}{\mathrm{T}}

\newcommand{\tg}{\mathrm{g}}
\newcommand{\tl}{\mathrm{l}}
\newcommand{\st}{\mathrm{st}}
\newlength{\dhatheight}

\usepackage{natbib} 
\usepackage{epsfig} 
\usepackage{epic} 
\usepackage{eepic} 
\usepackage{url} 
\usepackage{longtable} 
\usepackage{mathrsfs} 
\usepackage{multirow} 
\usepackage{bigstrut} 
\usepackage{graphicx} 
\usepackage{setspace} 
\usepackage{xspace} 
\usepackage{physics} 
\usepackage{booktabs} 
\usepackage{hyperref} 
\usepackage[noabbrev]{cleveref} 

\begin{document}
\title{Localization of Lindbladian Fermions}
\author{Foster Thompson$^1$}
\author{Yi Huang$^{1,2}$}
\author{Alex Kamenev$^{1,3}$}
\affiliation{$^1$School of Physics and Astronomy, University of Minnesota, Minneapolis, MN 55455, USA}
\affiliation{$^2$Condensed Matter Theory Center and Joint Quantum Institute, Department of Physics, University of Maryland, College Park, Maryland 20742, USA
}
\affiliation{$^3$William I. Fine Theoretical Physics Institute, University of Minnesota, Minneapolis, MN 55414, USA}

\begin{abstract}
We study a Lindbladian generalization of the Anderson model of localization that describes disordered free fermions coupled to a disordered environment.  
From finite size scaling of both eigenvalue statistics and participation ratio, we identify localization transitions in both the non-Hermitian Lindbladian spectrum, which governs transient relaxation dynamics, and in the Hermitian stationary state density matrix.  
These localization transitions occur at different critical values of Hamiltonian and dissipative disorder strength, implying the existence of atypical phases with a mixture of localized and delocalized features.  
We find this phenomenon is robust to changes to the value of the dissipative spectral gap.
\end{abstract}
\maketitle

\section{Introduction}
Disorder-driven phenomena are a central theme in condensed matter physics, with one of the most famous example being the Anderson model of localization \cite{Anderson1,Anderson2,Anderson3}.  As a successful phenomenological theory of metals and semiconductors, the Anderson model has demonstrated deep connections between disordered and generic quantum systems and random matrix theory \cite{LvlStatOriginal}.

While decades of progress has been made in disorder physics of closed systems, the effects of disorder in open quantum systems is a relatively new field of study, increasingly made accessible because of the recent availability of precisely controlled experiments that can be conducted on specific AMO systems.
One example of open quantum system with dissipation involves cold atoms confined within an optical lattice created by counter-propagating lasers, alongside the presence of an additional reservoir of cold atoms~\cite{Bloch:2005,LeBlanc:2007,Bloch:2008,Cazalilla_2014}. 
In this case, the gain is that atoms falling from the reservoir to the lattice, while the loss represents atoms escaping from the lattice trap.
Those experiments open up the possibility to discover new disorder-driven quantum phenomena.

There have been numerous recent theoretical explorations of the effects of disorder in Markovian quantum dynamics, governed by the Lindblad master equation \cite{OpenQM,QuantumNoise}.  Much recent effort has focused on random matrix models \cite{RMT1,RMT2,RMT3,RMT4,RMT5}, which demonstrated connections between generic Lindbladian spectra and non-Hermitian random matrix theory, a connection that has been proposed as both characteristic and defining of dissipative quantum chaos \cite{NonHermRatioDef,DissipativeChaos,SYK1,SYK2}.

In addition to the non-Hermitian Lindbladian spectrum, several works have examined the possibility of a nontrivial disorder-induced stationary state.  In \cite{RandomFerm,ManyBodyLindKeld}, quadratic Lindbladian fermions with all-to-all random matrix couplings were shown to undergo simultaneous transitions in both the spectrum and stationary state as a function of the number of decay channels.  Another set of examples is \cite{StationaryLocalization1,StationaryLocalization2}, which studied different types of localization transitions in the stationary states of disordered lattice fermions.

Divorced from the context of open quantum systems, various works have also developed the theory of non-Hermitian disorder-driven transitions, motivated for example by disordered photonics and random lasers \cite{RandomLaser1,RandomLaser2,RandomLaser3,TMEoriginal}.  A notable example is a non-Hermitian generalization of the Anderson model with complex-valued local potentials \cite{TMEoriginal,TME}, written to model disordered optical lattices with random gain and loss.  It was later shown that this model undergoes a non-Hermitian generalization Anderson localization in sufficiently high dimension \cite{Boris+Yi,Boris+Yi2,NonHermAndersonTransferMat,NonHermAndersonLevelStat2}.

In this work, we propose a Lindbladian generalization of the Anderson model which may act as a paradigmatic example of localization in open quantum systems.  This is achieved by introducing processes by which the system can gain and lose particles to and from a disordered environment.
To bridge the gap between existing theory on non-Hermitian Anderson transitions and disordered Lindbladians, we construct the model to have a spectrum determined by a non-Hermitian Anderson model and also to possess a nontrivial stationary state.
We study localization in our model using standard numerical techniques based on finite size scaling of both Hermitian and non-Hermitian random matrix level statistics and participation ratio (PR).
We find that both the spectrum and the stationary state localize and, strikingly, that these transitions occur at different disorder strengths.  This implies the existence of two interstitial phases between strong and weak disorder in which only one of the stationary state and transient modes are localized.

\section{Model and Results}\label{Sec:II}

Consider a gas of  fermions on a 3-dimensional square lattice with a total of $L^3$ sites.  Its many-body density matrix $\rho$ obeys the Lindblad equation:
\begin{equation}
\partial_t\rho=-i[\hat\fH,\rho]+\sum_v\bigg(\hat\fL_v\rho\hat\fL_v^\dagger-\frac{1}{2}\{\hat\fL_v^\dagger\hat\fL_v,\rho\}\bigg).
\end{equation}
The Hamiltonian is taken to be that of the standard Anderson model,
\begin{equation}
\hat\fH=
\sum_{\mathbf{r}}\varepsilon_\mathbf{r} \hat c_\mathbf{r}^\dagger\hat c_\mathbf{r} + \sum_{\langle\mathbf{r},\mathbf{r'}\rangle}\hat c^\dagger_\mathbf{r}\hat c_\mathbf{r'},
\end{equation}
where $\mathbf{r}$ is a lattice vector, and $\langle\mathbf{r},\mathbf{r'}\rangle$ denotes nearest-neighbor sites.  The onsite potential $\varepsilon_\mathbf{r}$ is random and independently distributed on each site, taken from a box distribution with width $W_\tR$, $\varepsilon_\mathbf{r}\in[-W_\tR/2,W_\tR/2]$.

To model coupling to a disordered environment, we allow each site to gain and lose particles at different independent random rates.  Each site may be thought of as coupled to two different baths, one pumps particles into the system and another acts as a sink to which particles can escape.  This is described by two sets of jump operators, corresponding respectively to loss and gain processes:
\begin{equation}
\hat\fL^{(\tl)}_\mathbf{r}=\mu_\mathbf{r}\hat c_\mathbf{r},\qquad\hat\fL^{(\tg)}_\mathbf{r}=\nu_\mathbf{r}\hat c_\mathbf{r}^\dagger.
\end{equation}
The dissipative couplings $\mu_\mathbf{r}$ and $\nu_\mathbf{r}$ are taken as random and independently distributed on each site.
The distribution of the dissipative couplings are chosen so that their squares are independently sampled from the  box distribution, $\mu_\mathbf{r}^2,\nu_\mathbf{r}^2\in[0,W_\tI]$.  This choice simplifies numerical computation and is not expected to modify universal features of the transitions.
The two sets of jump operators $\hat\fL^{(\mathrm{l,g})}_\mathbf{r}$ respectively correspond to the relative strengths of processes in which the system gains or loses particles.  This choice leads to both nontrivial transient relaxation dynamics and a nontrivial stationary state.

\subsection{Background}
Following the theory of quadratic Lindbladians based on either third quantization \cite{3rdQuant} or Keldysh techniques \cite{Kamenev2023,Sieberer2016,ManyBodyLindKeld}, one may introduce three single-particle matrices,
\begin{subequations}\label{HQD}
\begin{equation}
(H_0)_{\mathbf{r}\mathbf{r'}}= \varepsilon_\mathbf{r}
\, \delta_{\mathbf{r}\mathbf{r'}} + \delta_{\langle\mathbf{r},\mathbf{r'}\rangle},
\end{equation}
\begin{equation}
Q_{\mathbf{r}\mathbf{r'}}=\frac{1}{2}(\mu^2_\mathbf{r}+\nu^2_\mathbf{r})\delta_{\mathbf{r}\mathbf{r'}},
\end{equation}
\begin{equation}
D_{\mathbf{r}\mathbf{r'}}=(\mu^2_\mathbf{r}-\nu^2_\mathbf{r})\delta_{\mathbf{r}\mathbf{r'}}.
\end{equation}
\end{subequations}
The single-particle eigenvalues of the Lindbladian are the equivalent to the eigenvalues of the non-Hermitian dynamic matrix $H=H_0-iQ$.  The corresponding eigenvalues are the single-particle transient modes of the many-body dynamics, governing relaxation to the stationary state.

In contrast, the (unique) stationary state density matrix $\rho_\st$ is a Gaussian state that can be expressed using an effective Hermitian Hamiltonian,
\begin{equation}
\rho_\st\propto\exp\Bigg\{-\sum_{\mathbf{r},\mathbf{r'}}
\hat c^\dagger_\mathbf{r}\,
(H_\st)_{\mathbf{r}\mathbf{r'}}\,\hat c_\mathbf{r'}\Bigg\}.
\end{equation}
The matrix $H_\st$ may be determined from the stationary Keldysh distribution function $F_\st$, which in turn solves a matrix Lyapunov equation \cite{ManyBodyLindKeld},
\begin{subequations}\label{defF}
\begin{equation}
F_\st=\tanh(H_\st/2);
\end{equation}
\begin{equation}\label{Lyapunov}
i(HF_\st-F_\st H^\dagger)=D .
\end{equation}
\end{subequations}
The stationary state $\rho_\st$ is a statistical superposition of free fermion states with different occupation probabilities.  The eigenvectors of $H_\st$ determine the single particle states and the eigenvalues specify their corresponding stationary occupation numbers.  The uniqueness of the stationary state is a consequence of the absence of conserved quantities in the model. One cannot specify a temperature or chemical potential because neither energy nor particle number is conserved.

Thus two distinct objects, the dynamic matrix $H$ and the stationary Hamiltonian $H_\st$, are needed to fully characterize the theory.  This should be contrasted with coherent models of disordered free fermions, for which all features of the model can be deduced from the properties of the single-particle Hamiltonian alone for a given temperature and chemical potential.

In the context of the model considered here this presents two distinct random matrix problems, one non-Hermitian and one Hermitian.  The dynamic matrix $H$ is similar to a non-Hermitian Anderson model for example in Refs.~\cite{TMEoriginal,TME}, but the sampling of the imaginary on-site terms chosen to ensure the spectrum is supported only below the real axis.  This difference can be expected to manifest only in non-universal features of the model, so one should anticipate a non-Hermitian Anderson transition in the AI$^{\dagger}$ symmetry class \cite{NonHermSymClasses}.

In contrast, the stationary Hamiltonian is specified indirectly through the constraint imposed by the Lyapunov equation eq.~(\ref{Lyapunov}).  This specifies an unusual random matrix problem,  which to our knowledge has not been previously studied.  The entries of the matrix $H_\st$ may be complex, placing it in the unitary symmetry class A.  This is true in spite of the fact that all three single-particle matrices $H_0$, $Q$, and $D$ are all orthogonal matrices with purely real entries.  This is a consequence of the dynamic lack of time reversal symmetry: the Lindbladian dynamics describes the irreversible process of a system relaxing to a unique stationary state while in contact with its environment.  One may thus expect a localization transition within the unitary universality class.

\subsection{Summary of Main Results}
\begin{figure}
    \includegraphics[width=1\linewidth]{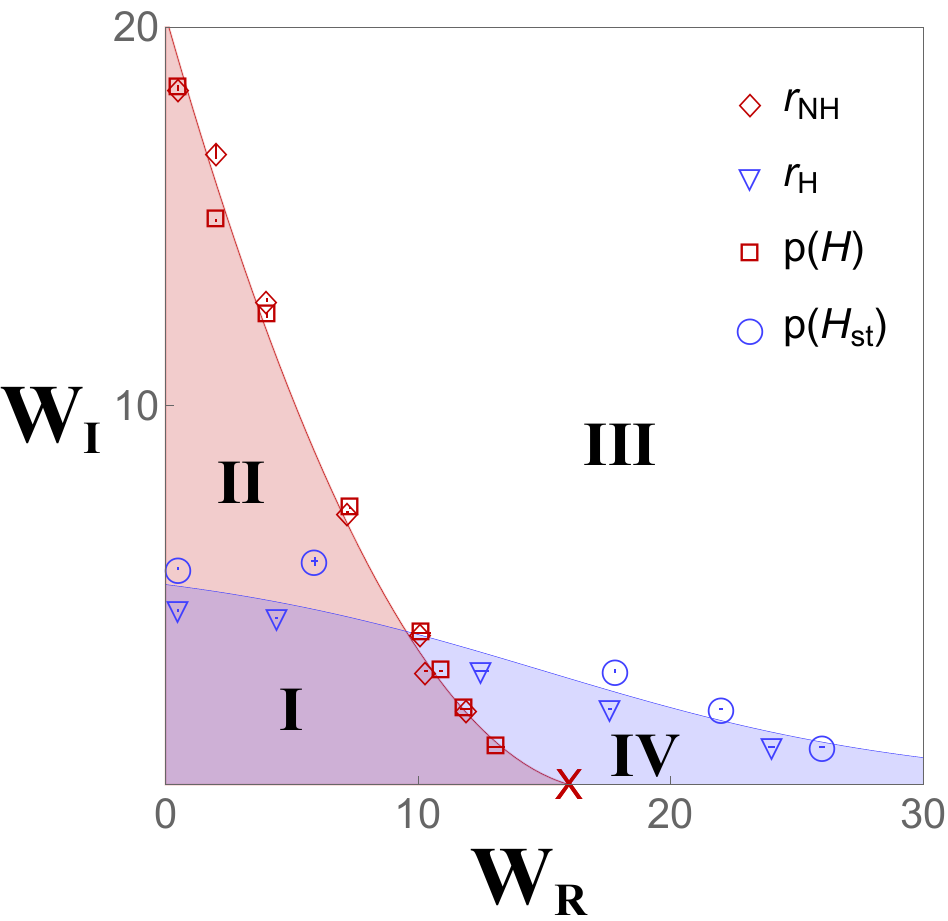}
    \caption{Effective phase diagram of the 3d model.  Dark red and light blue markers denote numerically determined critical disorder strengths $(W_\tR^*,W_\tI^*)$ (see section \ref{Sec:III}) for the dynamic matrix, $H$, and the stationary state, $H_\st$, respectively; lines through their centers denote numerical uncertainties.  The triangle and diamond shape markers denote values determined from ratio statistics and circle and square markers -- from the participation ratio.  The dark red X on the horizontal axis denotes the critical point of the Hermitian Anderson model.  Solid lines illustrate schematic phase boundaries consistent with our numerical estimates which separate four phases, labeled with numerals.  Phases I and II have delocalized $H$ and respectively delocalized and localized $H_\st$; phases III and IV have localized $H$ and respectively localized and  delocalized $H_\st$.}
\label{fig:PhaseDiagram}
\end{figure}

The model has two parameters, $W_\tR$ and $W_\tI$, respectively the Hamiltonian and dissipative disorder strengths.  The limit of strong disorder of either type causes both the dynamic matrix $H$ and the stationary state effective Hamiltonian $H_\st$ to localize.  These transitions occur at different critical disorder strengths, resulting in four distinct phases, see fig.~\ref{fig:PhaseDiagram}.  Two phases have a mixture of localized and delocalized features: the stationary state is a statistical mixture of either delocalized or localized states, while the relaxation to this state occurs through the excitation of localized or delocalized transient modes, respectively.  This phenomenon has no analogue in equilibrium, where the state and dynamics are canonically specified by the same Hamiltonian.  It is possible in the present context because fluctuations in the stationary state and relaxation dynamics have no fixed relation (no FDT) in the far-from-equilibrium dynamics studied here.

We further find that the existence of these phases is not strongly contingent on the dissipative spectral gap -- the smallest imaginary part of all nonzero Lindbladian eigenvalues.
It is instead the property of the spectrum near the center of the support of the single particle matrices, where their density of states is large, that governs the transition, similar to the conventional Hermitian Anderson localization.
This suggests that these localization phenomena are conceptually unrelated to the rate of approach to the stationary state, and may occur in systems with or without a finite dissipative gap.
For a more detailed discussion and justification, see section \ref{Sec:IIIC}.

We note several additional features of the phase diagram before discussing details of the transition in section \ref{Sec:III}.  The location of the critical line of $H$ is is relatively symmetric with respect to $W_\tR$ vs $W_\tI$.  This is expected, as similar results hold for the non-Hermitian Anderson model with different real and imaginary disorder strengths \cite{NonHermAndersonTransferMat}.  The slight asymmetry in our model comes from the difference in the sampling of real and imaginary on-site potentials.  Qualitatively both play a similar role in driving the localization transition of $H$.  The dissipationless limit $W_\tI\to0$ corresponds to $H\to H_0$.  The critical line for $H$ intersects the horizontal axis at $W_\tR\simeq16$, the known value for Hermitian Anderson localization on a $d=3$ square lattice with nearest-neighbor hopping \cite{WcValue}.

In contrast, the critical line of $H_\st$ is comparatively quite flat in the horizontal direction.  The location of the critical point depends only very weakly on $W_\tR$ until its value becomes very large.  This lack of sensitivity to the Hamiltonian disorder strength suggests that stationary state localization is driven almost exclusively by competition between the coherent hopping and incoherent gain and loss processes.
For large $W_\tR$, the phase boundary tends toward the horizontal axis.  In the range of small $W_\tI$ and large $W_\tR$, denoted phase IV in fig.~(\ref{fig:PhaseDiagram}), $H_\st$ is delocalized,  despite the single-particle Hamiltonian $H_0$ being strongly localized.
The ultimate fate of phase IV for very large $W_\tR$ is difficult to resolve numerically.  It is unclear if it eventually terminates in a second critical point on the horizontal axis, or if the stationary state remains delocalized for any large but finite $W_\tR$.  
The Lyapunov equation becomes ill-defined in the dissipationless limit, $W_\tI=0$, reducing to a commutator with $H_0$ and thus lacking a unique solution.  The different phases for $H_\st$ are meaningful only for a finite $W_\tI>0$. 

\section{Numerics}\label{Sec:III}
We perform a finite-size scaling analysis based on both eigenvalue level statistics and PR on both $H$ and $H_\st$.  We also examine the localization profile of individual eigenvectors of both through the PR and estimate critical exponents of the transitions.  We conclude with a discussion of the role of the dissipative gap.

\subsection{Transient Modes}\label{Sec:IIIA}
Localization of the dynamic matrix $H$ can be detected through non-Hermitian eigenvalue statistics:
letting $s_{1,2}(\epsilon)$ denote the geometric distance in the complex plane between an $H$ eigenvalue $\epsilon$ and its nearest and next-nearest neighbors resp., the non-Hermitian level spacing ratio is defined as \cite{NonHermRatioDef},
\begin{equation}
r_\mathrm{NH}(W_\tR,W_\tI)=\ev{s_1(\epsilon)/s_2(\epsilon)},
\end{equation}
where $\ev{\dots}$ is both a disorder average and a sum over all $\epsilon$ in a finite-size window centered in the middle of the spectrum of $H$.  The window is defined as the middle quarter of eigenvalues sorted by the magnitude of both their real and imaginary parts.  This quantity has the limiting values $r_\mathrm{NH}=0.67$ for uncorrelated eigenvalues in a localized system and $r_\mathrm{NH}=0.72$ for the symmetry class AI$^\dagger$ in a delocalized system \cite{NonHermRatioDef}.

\begin{figure}
    \includegraphics[width=1\linewidth]{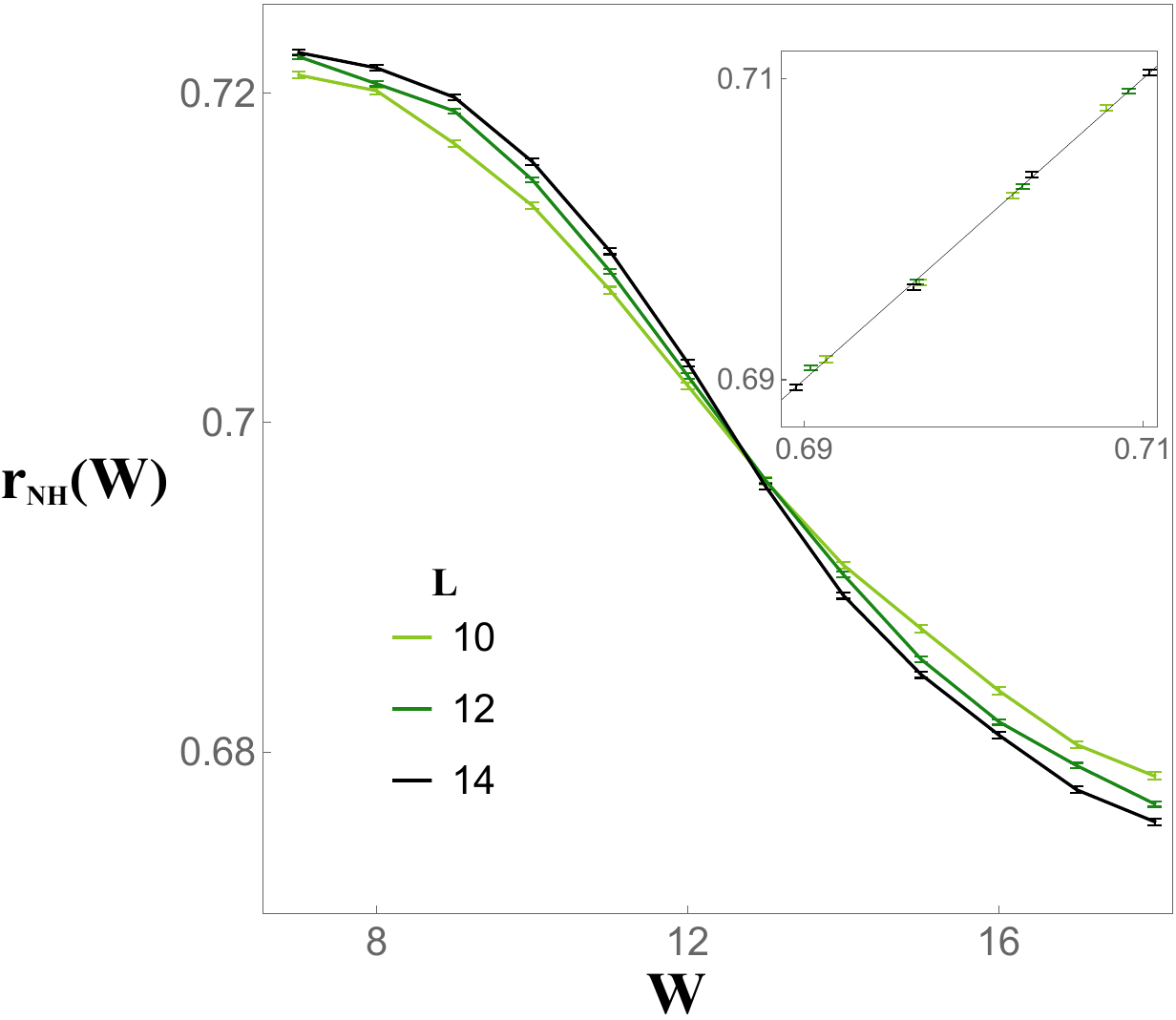}
    \includegraphics[width=1\linewidth]{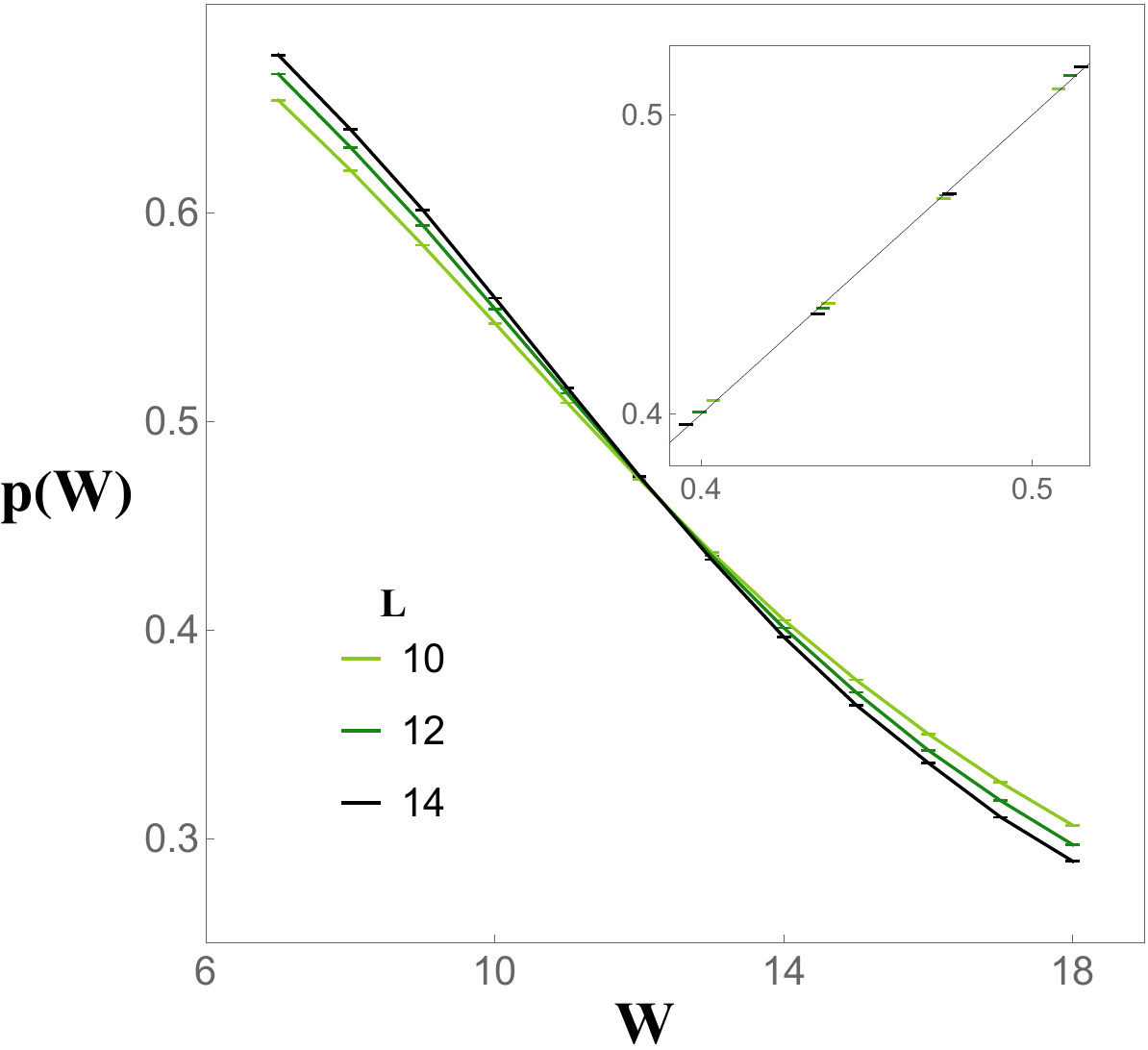}
    \caption{Examples of finite size scaling of level statistics and PR for $H$, with fixed $W_\tR=4$.  The horizontal axis $W$ on both plots is $W_\tI$; the vertical axes show $r_\mathrm{NH}$ and $p$ with uncertainties.  The crossing points are $W_\tI=12.77\pm0.02$ and $12.36\pm0.06$ respectively, giving critical points in the $(W_\tR,W_\tI)$ plane of $(4,12.77)$ and $(4,12.36)$.  The inserts show computed values of $r_\mathrm{NH}(W)$ and $p(W)$ using the scaling forms given in eq.s~(\ref{rscaling}) and (\ref{pscaling}) with parameters determined from fitting vs. their numerically determined values; a solid line with unit slope is shown for reference.  The estimated critical exponent is $\nu=1.44\pm0.03$ from level statistics and $\nu=1.23\pm0.02$ from PR.}
    \label{fig:NonHermData}
\end{figure}

By either fixing one of $W_{\tR,\tI}$ and varying the other or by imposing a constraint between the two, the curves of $r_\mathrm{NH}(W)$ for different system sizes $L$ all intersect at a critical point $(W_\tI,W_\tR)=(W_\tI^*,W_\tR^*)$, Fig.~\ref{fig:NonHermData}.  The precise location of the critical point is extracted by fitting data to the scaling form in the vicinity of the transitions,
\begin{equation}\label{rscaling}
r_\mathrm{NH}(W)\simeq r_\mathrm{NH}^*-r_\mathrm{NH}^{(1)}L^{1/\nu}(W-W^*),
\end{equation}
where $r_\mathrm{NH}^*$ is the value of $r_\mathrm{NH}$ at the critical point, $r_\mathrm{NH}^{(1)}$ is a constant of proportionality, and $\nu$ is the critical exponent of the correlation length.

\begin{figure}
    \includegraphics[width=1\linewidth]{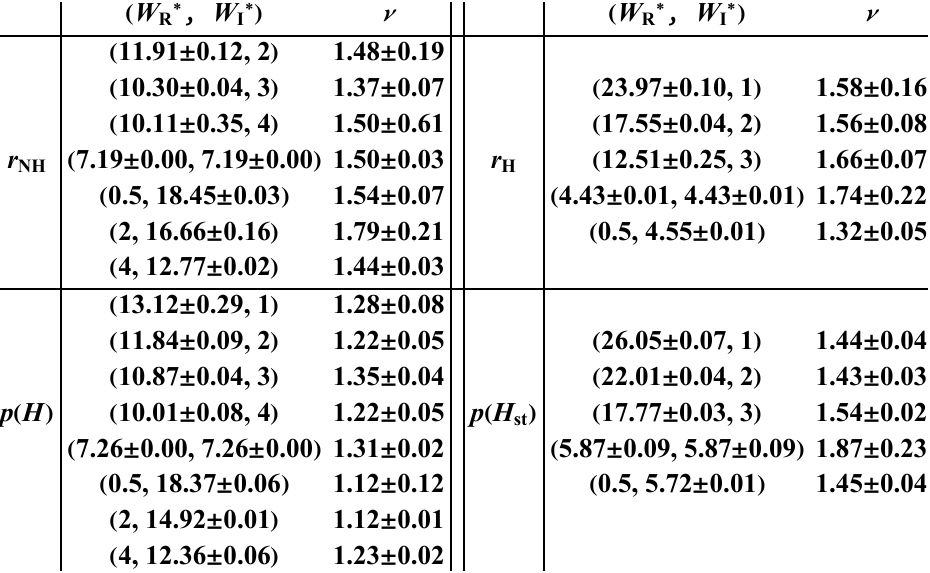}
    \caption{Fitted values of the localization length critical exponent $\nu$ compared to critical disorder strengths $(W_\tR^*,W_\tI^*)$ obtained from finite size scaling analysis of both ratio statistics and PR for both $H$ and $H_\st$, shown with estimated numerical uncertainty.}
\label{fig:Table}
\end{figure}

We verify the existence of the transition by studying spatial profile of the eigenvectors of $H$.  Using the right eigenvector of $H$, given by $H\ket{\epsilon}=\epsilon\ket{\epsilon}$, one may define the participation ratio, similarly to that in Hermitian models \cite{PRdef1,PRdef2}:
\begin{equation}
\mathrm{PR}(\epsilon)=\qty[\sum_\mathbf{r}\frac{|\ev{\mathbf r|\epsilon}|^4}{|\braket{\epsilon}|^2}]^{-1},
\end{equation}
where $\ket{\mathbf r}$ is the spatial basis on the lattice.  Focusing only on right eigenvectors of $H$ is justified because $H=H^\tT$ and therefore the left eigenvectors are the complex conjugates of right ones thus having the same PR.  To perform the scaling analysis, we introduce the following quantity:
\begin{equation}\label{PR}
p(W_\tR,W_\tI)=\ev{\frac{\log(\mathrm{PR})}{\log(L^d)}},
\end{equation}
where $\ev{\dots}$ is the same disorder plus spectral average as used in $r_\mathrm{NH}$, described above.  Near the transition this quantity has the scaling form,  \cite{PRscaling1,PRscaling2,PRscaling3},
\begin{equation}\label{pscaling}
p(W)\simeq p^*-p^{(1)}\frac{L^{1/\nu}}{\log(L)}(W-W^*),
\end{equation}
where as in Eq.~(\ref{rscaling}) $W$ is either of $W_{\tR,\tI}$ and $\nu$ is the critical exponent; $p^*$ is the value of $p$ at the critical point, $p^{(1)}$ is a constant of proportionality.  
Examples of the finite-size scaling of both ratio statistics and PR for are shown in Fig.~\ref{fig:NonHermData}.
This fitting also enables estimation of critical exponent $\nu$, see Fig.~\ref{fig:Table}.  The values found using the ratio statistics are most closely consistent with $\nu\simeq1.5$, obtained for $d=3$ in Ref.~\cite{Boris+Yi}; those obtained from PR are closer to the the estimate of $\nu\simeq1.19$ found in Ref.~\cite{NonHermAndersonTransferMat}.

\begin{figure}
    \includegraphics[width=1\linewidth]{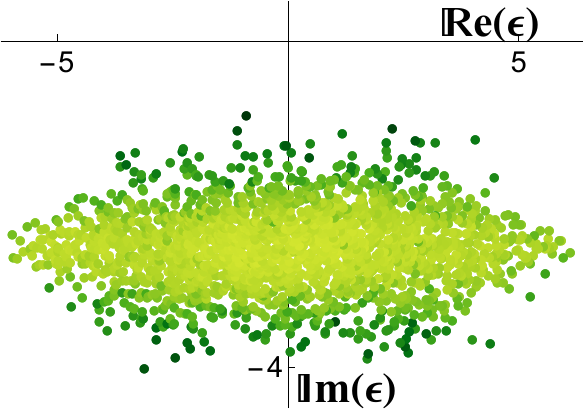}
    \includegraphics[width=1\linewidth]{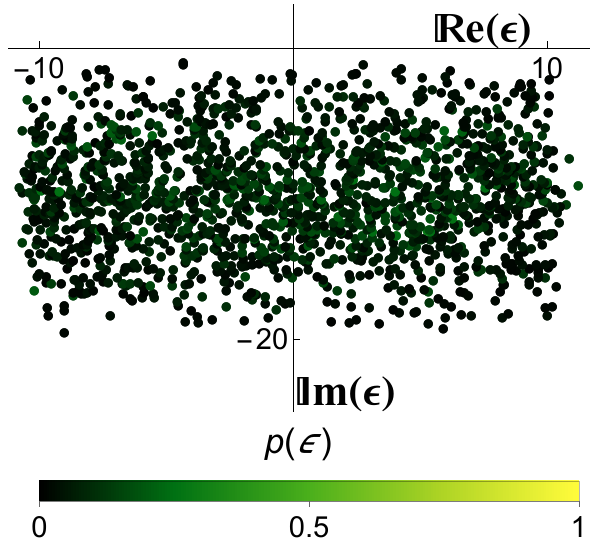}
    \caption{Complex spectra of $H$ with $L=12$ and $(W_\tR,W_\tI)=(5,5)$ above and $(W_\tR,W_\tI)=(20,20)$ below.  Colors indicate the value of $p(\epsilon)=\log(PR(\epsilon))/\log(L^3)$, so that more localized eigenmodes are shown appear darker.}
\label{fig:PRexeNonherm}
\end{figure}

We also examine the PR of each eigenvector compared to its location in the complex plane, shown in Fig.~\ref{fig:PRexeNonherm}.  The result is similar to what was observed in Ref.~\cite{TME}.  For a localized $H$, eigenvalues are nearly spread over the entire square-shaped region with width $W_\tR$ and height $W_\tI$, with some concentration near the vertical center due to the choice of sampling.  For $H$ delocalized, the majority of eigenvalues are concentrated in the middle of the density of states and are delocalized. However, near the boundary of the support region there is a small number of  localized states on all sides.  This is the complex-plane analogue of the mobility edge in the conventional Anderson model.  In the present Lindbladian context, it has an additional significance: the eigenvalues closest to the real axis correspond to modes with the smallest decay rate.  Since they fall on the edge of the spectrum, it implies that the longest living transient modes tend to be localized.

\subsection{Stationary State}\label{Sec:IIIB}
\begin{figure}
    \includegraphics[width=1\linewidth]{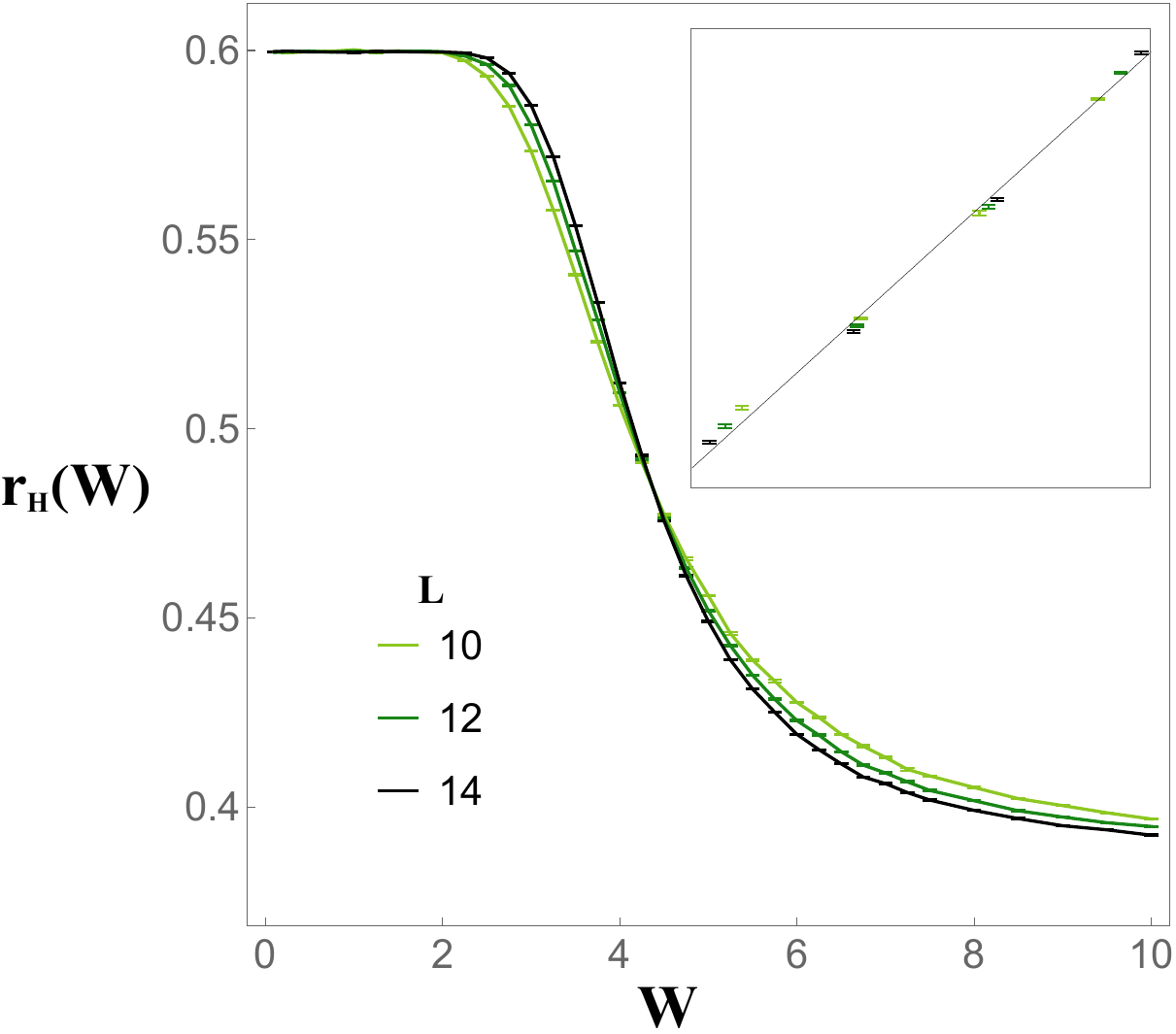}
    \includegraphics[width=1\linewidth]{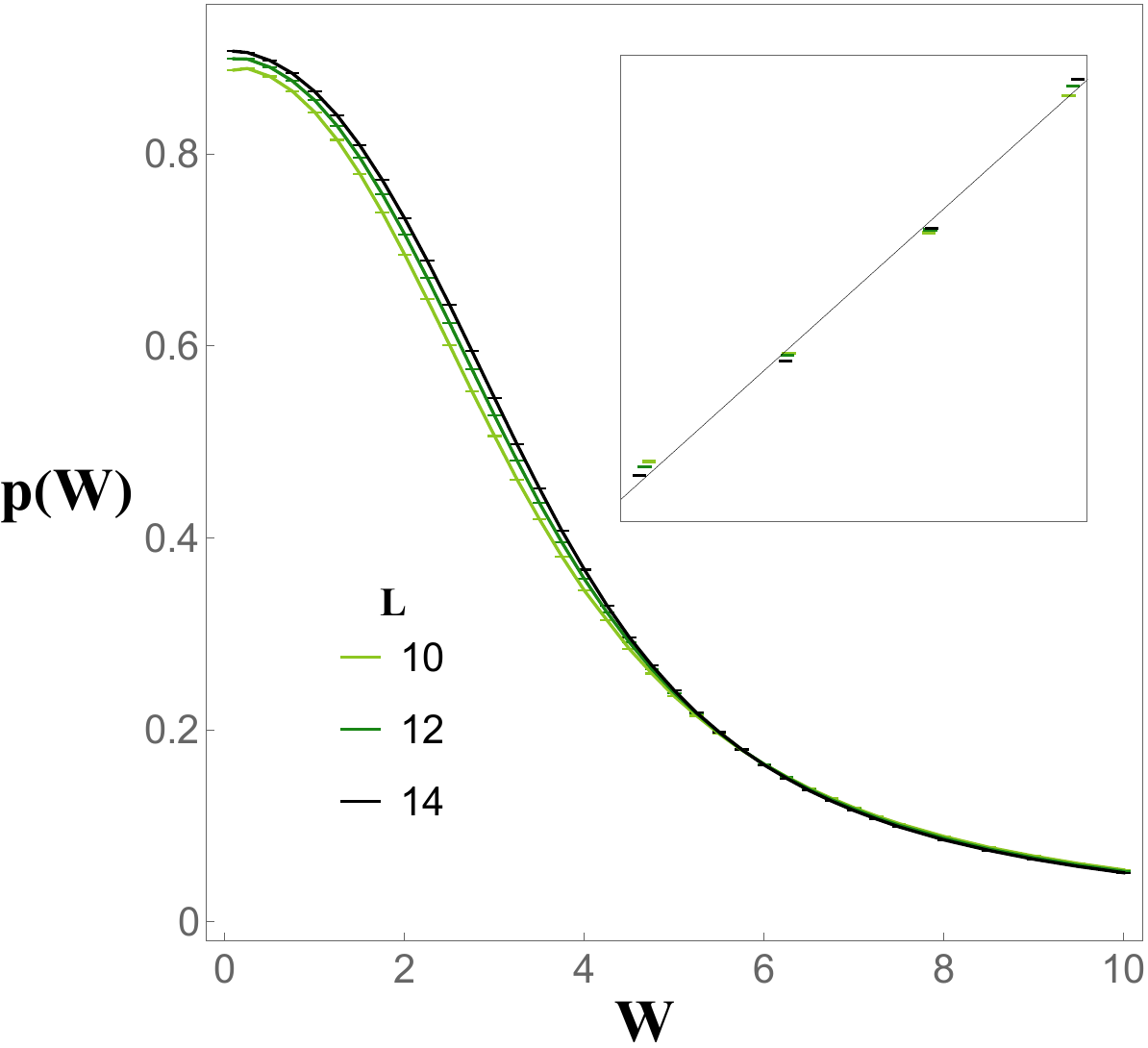}
    \caption{Examples of finite size scaling of level statistics and PR for $H_0$, with fixed $W_\tR=W_\tI$.  The horizontal axis shows varying values of $W=W_{\tR,\tI}$; the vertical axes show $r_\mathrm{H}$ and $p$ with uncertainties.  The crossing points are $W=4.43\pm0.01$ and $5.87\pm0.09$ respectively, giving critical points $(4.43,4.43)$ and $(5.87,5.87)$.  The inserts show computed values of $r_\mathrm{NH}(W)$ and $p(W)$ using the scaling forms given in Eqs.~(\ref{rscaling}) and (\ref{pscaling}) with parameters determined from fitting vs. their numerically determined values; a solid line with unit slope is shown for reference.  The estimated critical exponent is $\nu=1.74\pm0.22$ from level statistics and $\nu=1.87\pm0.24$ from PR.}
    \label{fig:HermData}
\end{figure}

The stationary state effective Hamiltonian $H_\st$ undergoes a localization transition in the unitary symmetry class A, which is shown for a fixed value of $W_\tI$ in Fig.~\ref{fig:HermData}.  To detect this transition, we again use both ratio statistics and participation ratio.  Ratio statistics for Hermitian matrices is defined by letting $\beta_n$ denote the $n$th eigenvalue of $H_\st$ ordered from smallest to largest and $s_n=\beta_{n+1}-\beta_n$.  The level spacing for Hermitian matrices is, \cite{HermRatioDef,HermRatioDef2},
\begin{equation}
r_\mathrm{H}(W_\tR,W_\tI)=\ev{\min\bigg\{\frac{s_n}{s_{n-1}},\frac{s_{n-1}}{s_n}\bigg\}},
\end{equation}
where $\ev{\dots}$ is the disorder average and an average over a fraction of eigenvalues in the center of the spectrum of $H_\st$, defined using the middle half of eigenvalues sorted by magnitude.
In the localized phase, eigenvalues are uncorrelated and $r_\mathrm{H}=0.39$; in the delocalized phase, correlations match the GUE value of $r_\mathrm{H}=0.60$ \cite{HermRatioDef}.

Analogous to the previous section, we identify a transition by fixing one of either $W_{\tR,\tI}$ and performing a finite size scaling analysis by varying the other.  The scaling form of $r_\mathrm{H}$ is the same as that of $r_\mathrm{NH}$ shown in Eq.~(\ref{rscaling}).  The existence of the transition is confirmed using PR, which is defined the same way as in Eq.~(\ref{PR}).  As with $H$, we find a critical line of transition values of $(W_\tR^*,W_\tI^*)$ which separate localized and delocalized phases of $H_\st$, shown in Fig.~\ref{fig:PhaseDiagram}.
The scaling exponent estimates predicted from both ratio statistics are close to known results for example in Refs.~\cite{GUEnu1,GUEnu2} in which the $d=3$ Anderson transition in the symmetry class A was found to have a critical exponent which is close to $\nu\simeq1.44$; see Fig.~\ref{fig:Table}.

\begin{figure}
    \includegraphics[width=.49\linewidth]{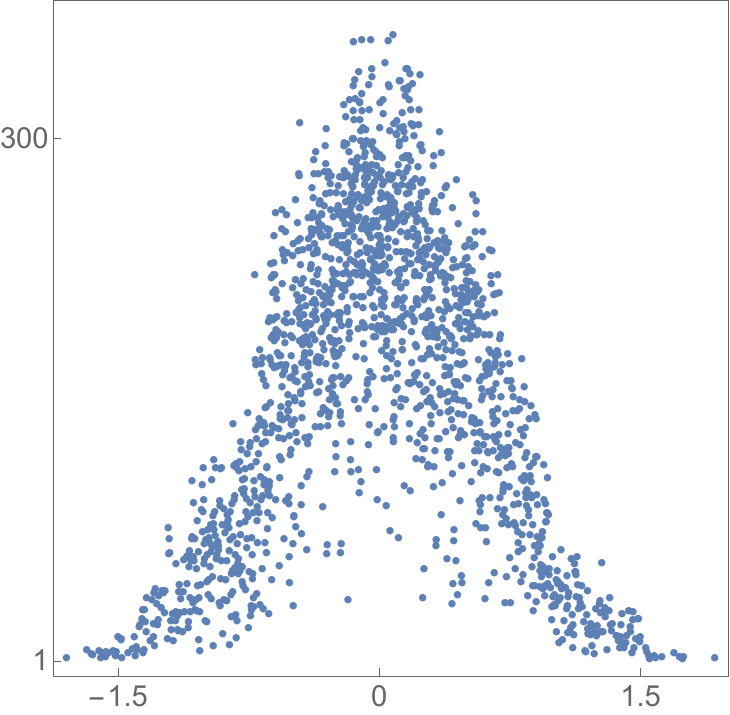}
    \includegraphics[width=.49\linewidth]{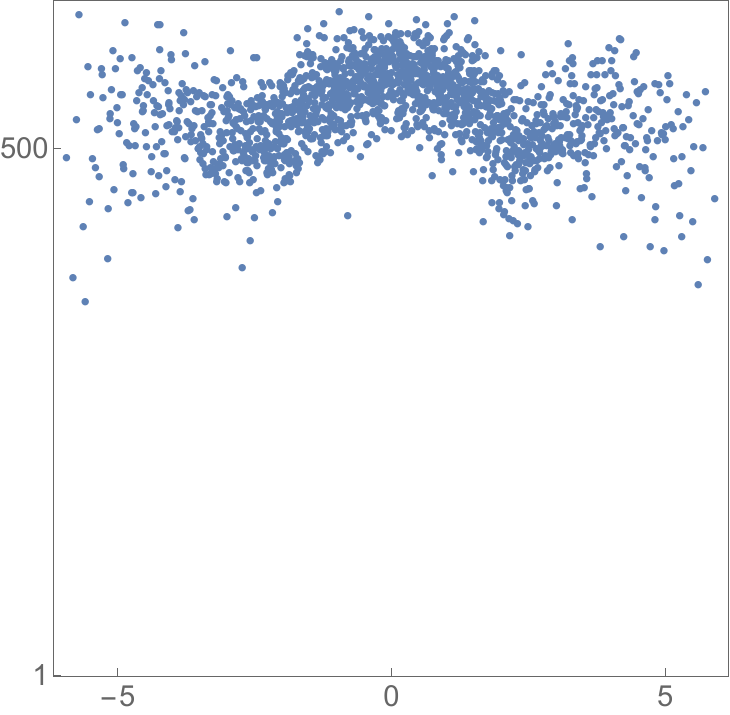}
    \includegraphics[width=.49\linewidth]{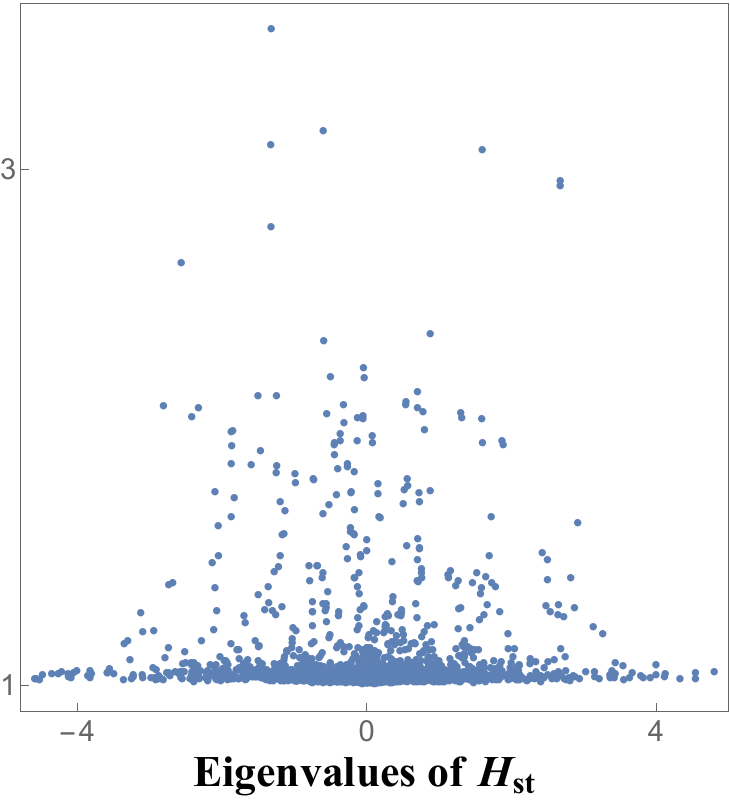}
    \includegraphics[width=.49\linewidth]{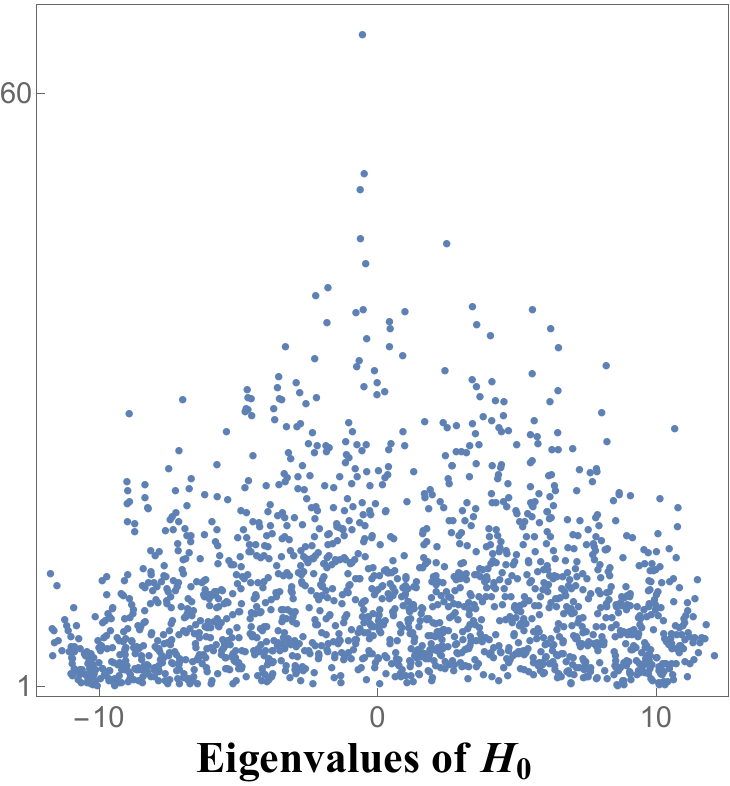}
    \caption{Participation ratios (vertical axes on all plots) for the eigenvectors of the steady state effective Hamiltonian $H_\st$ (left column) and the corresponding Anderson Hamiltonian $H_0$ (right column) from the same disorder realization, shown for $L=12$.  The bottom row shows $(W_\tR,W_\tI)=(2,2)$ where both are in the delocal phase; the top row shows $(W_\tR,W_\tI)=(20,20)$ where both are localized.
    }
    \label{fig:PRexeHerm}
\end{figure}

We also observe that the profile of the PR has qualitative differences from that of the conventional Anderson model, see Fig.~\ref{fig:PRexeHerm}.  The overall shape of the distribution of the PR over the spectrum of $H_\st$ is substantially more pronounced than the those of the corresponding $H_0$.  We notice that in the delocalized phase, the distribution exhibits a clear mobility edge, showing a prevalence of delocalized states with large PR at the center of its density of states and states that are more localized with smaller PR near the edges.  The bandwidth of $H_\st$ is quite small compared to that of $H_0$, especially in the delocal phase.

These numerical findings justify our focus on the eigenstates of $H_\st$ near the center of its density of states to identify the transition.
For delocal stationary states the bandwidth of $H_\st$ is small, implying that all eigenstates contribute to expectation values with roughly equal probability.  The density of states near the middle is much larger than the edge, so observables are dominated by delocalized features.  
For local stationary states, all states are localized, so expectation values always show localized features.

\subsection{Dissipative Gap}\label{Sec:IIIC}
The dissipative gap determines the slowest rate of approach towards the stationary state from an arbitrary initial state.  Formally, it is the minimum of the imaginary part of the Lindbladian spectrum.  Systems with a finite dissipative decay towards the stationary state exponentially with the timescale given by the inverse gap.  Systems without the gap, in contrast, can decay algebraically.

For the quadratic Lindbladian dynamics considered here, the many-body dissipative gap is the same as that of $H$.  Any particular disorder realization exhibits a finite dissipative gap (though this may be lost in the limit $L\to\infty$).
In particular, when $H$ is strongly localized the eigenvalues of $H$ are determined almost exactly by local $\varepsilon_\mathbf{r}$, $\mu_\mathbf{r}$, and $\nu_\mathbf{r}$ and thus fill the entire square-shaped region in the complex plane bound by the real and imaginary intervals $[-W_\tR/2,W_\tR/2]$ and $[0,W_\tI]$.  Because both $\mu_\mathbf{r}$ and $\nu_\mathbf{r}$ can be arbitrarily small, eigenvalues can be arbitrarily close to the real axis, leading to a vanishing dissipative gap.

While this is true of the model as defined above, we argue that this is not an essential feature of either type localization observed here.  A modification to the jump operators can introduce a finite dissipative gap even in the strongly localized regime of either or both of $H$ and $H_\st$.
To show this, we consider an additional set of non-random gain and loss channels encoded by two additional sets of jump operators:
\begin{equation}
\hat\fL_\mathbf{r}^{(\mathbf{l}')}=\sqrt\kappa\, \hat c_\mathbf{r},\qquad\hat\fL_\mathbf{r}^{(\mathbf{g}')}=\sqrt\kappa\, \hat c_\mathbf{r}^\dagger,
\end{equation}
where $\kappa$ is a site-independent constant.  This addition modifies the single particle matrix $Q$ of Eq.~(\ref{HQD}),
\begin{subequations}
\begin{equation}
Q_{\mathbf{r}\mathbf{r'}}=\frac{1}{2}\,(\mu^2_\mathbf{r}+\nu^2_\mathbf{r}+2\kappa)\,\delta_{\mathbf{r}\mathbf{r'}}.
\end{equation}
\end{subequations}
The other two single-particle matrices are unchanged.  Note that the same could have been achieved by modifying the distribution of $\mu_\mathbf{r}$ and $\nu_\mathbf{r}$ so that their squares are uniformly distributed on the shifted interval $[\kappa,W_\tI+\kappa]$ instead of introducing new jump operators.

\begin{figure}
    \includegraphics[width=1\linewidth]{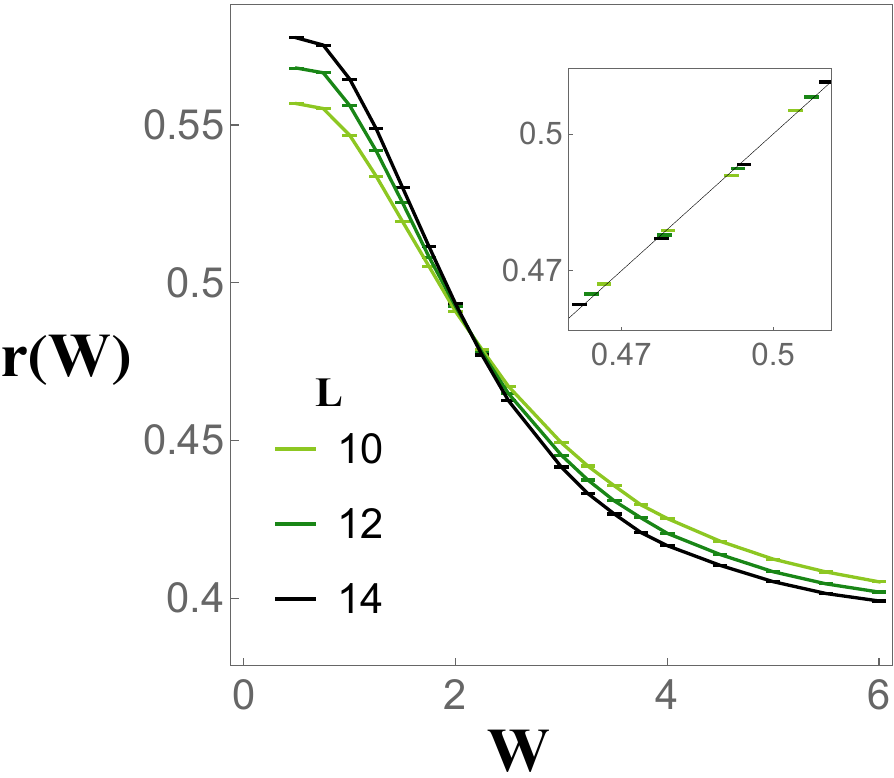}
    \includegraphics[width=1\linewidth]{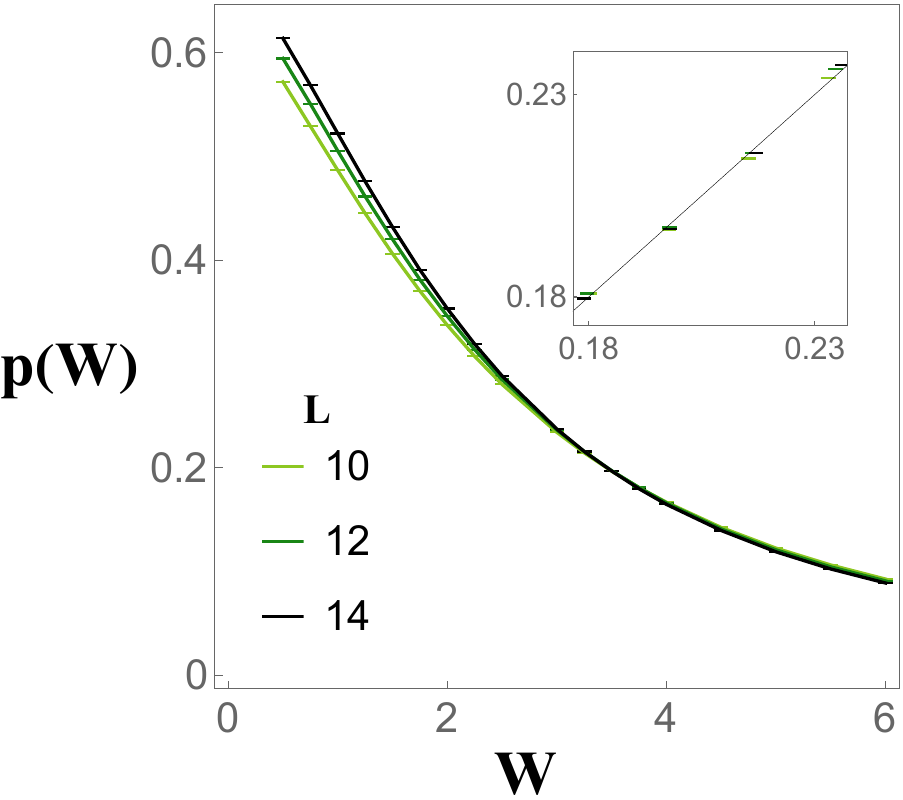}
    \caption{Finite size scaling of level statistics and PR for $H_\st$ with $\kappa=1$ and $W_\tR=W_\tI$, compare to fig.~(\ref{fig:HermData}).  The horizontal axis shows varying values of $W=W_{\tR,\tI}$ and the vertical axes show $r_\mathrm{H}$ and $p$ with uncertainties.  The crossing points are $W_\tI=3.53\pm0.03$ and $2.17\pm0.002$ respectively, giving critical points $(3.53,3.53)$ and $(2.17,2.17)$.  The inserts show computed values of $r_\mathrm{H}(W)$ and $p(W)$ using the scaling forms given in eq.s~(\ref{rscaling}) and (\ref{pscaling}) with parameters determined from fitting vs. their numerically determined values; a solid line with unit slope is shown for reference.
    The estimated critical exponent is $\nu=1.32\pm0.05$ from level statistics and $\nu=1.57\pm0.12$ from PR.}
    \label{fig:GapData}
\end{figure}

This translates the entire spectrum of $H$ away from the real axis by a distance $\kappa$ in the complex plane, so the dissipative gap is at least as large as $\kappa$ regardless of the strength of the disorder.
This change does not modify any other details of the eigenvalues or eigenvectors.  The non-Hermitian Anderson transition is thus unaffected and the critical line for $H$ in the $(W_\tR,W_\tI)$ plane, depicted in Fig.~\ref{fig:PhaseDiagram}, is unchanged.
The effects on the stationary state are less obvious, but we numerically verified that its localization transition remains intact and the location of the critical points are not affected for small but finite $\kappa$.  We find that the location of critical points generally move toward the origin of the $(W_\tR,W_\tI)$ with increasing $\kappa$ but remained finite, see Fig.~(\ref{fig:GapData}).

\section{Conclusion}
We have proposed a simple Lindbladian model of localization in an open system of disordered fermions.  By studying both eigenvalue statistics and and participation ratio, we found that both the transient modes and the stationary state undergo localization transitions at sufficiently strong disorder of either the Hamiltonian or dissipative type.  We established a schematic phase diagram of our model and estimated the localization length critical exponent of the two transitions, which generally agreed with known results for their respective universality classes along the entire phase boundary.

Surprisingly, we found that the stationary state is much more sensitive to the dissipative disorder, while the transient mode spectrum is affected by both types of disorder in an approximately symmetric way.
As a consequence, localization of the transient modes and stationary state occur at different critical disorder strengths.
This results in four distinct phases: two conventional phases corresponding to weak and strong disorder of both types in which all features are delocalized or localized respectively, and two unconventional phases in which only one of the stationary state or transient modes is localized, while the other remains delocalized.
We showed that the phases persist independently on the presence of the dissipative spectral gap.

One may speculate on the observable consequences of these new types of localization.  Following Ref.~\cite{ManyBodyLindKeld}, expectation values of observables and their equal-time higher correlation functions depend only on $H_\st$.  The dynamic matrix $H$ determines observables' non-stationary properties, such as linear response features or quenches from initially non-stationary states.
Measurements of local quantities cannot detect a sharp transition between localized and delocalized phases.
In the Hermitian Anderson model, localization is signalled by vanishing conductivity.  Such a metric is unsuitable in the model studied here because it has no conserved quantities and hence no transport features.
We leave the question of an appropriate experimental signature of localization in this model to future work.

A possible alternative way forward is the construction of a more complicated Lindbladian model with similar localization phenomena that also possesses conserved quantities.  In such a theory, transport features could be a signal of localization.  This would necessitate going beyond the single-particle description, as the inclusion of conserved quantities requires jump operators that are at least quadratic in fermion creation operators.
It is an open question as to whether or not mixed localized and delocalized features could be realized in a disordered many-body Lindbladians, but it is beyond the scope of the this manuscript.


\section{Acknowledgments} 

We are indebted to Boris Shklovskii for  stimulating discussions. The work was supported by NSF grants DMR-2037654 and DMR-2338819. 

\newpage
\bibliographystyle{apsrev4-2}
\bibliography{reference}

\end{document}